\documentclass[prd,nofootinbib,preprint,superscriptaddress]{revtex4-1}
\usepackage{comment}
\usepackage{amsmath, amssymb, amsthm, graphicx, epsfig, fancyhdr,epsfig}
\usepackage{physics}
\usepackage{braket}
\usepackage{tikz-feynman}
\tikzfeynmanset{compat=1.1.0}
\usepackage{hyperref}
\usepackage{tikzsymbols}

\usepackage{float}
\usepackage{xcolor}

\newcommand{\be}{\begin{equation}}
\newcommand{\ee}{\end{equation}}
\newcommand{\bea}{\begin{eqnarray}}
\newcommand{\eea}{\end{eqnarray}}

\DeclareRobustCommand{\orcidicon}{
	\begin{tikzpicture}
	\draw[lime, fill=lime] (0,0) 
	circle [radius=0.2] 
	node[white] {{\fontfamily{qag}\selectfont \tiny ID}};
	\draw[white, fill=white] (-0.0625,0.095) 
	circle [radius=0.007];
	\end{tikzpicture}
	\hspace{-2mm}
}

\foreach \x in {A, ..., Z}{\expandafter\xdef\csname orcid\x\endcsname{\noexpand\href{https://orcid.org/\csname orcidauthor\x\endcsname}
			{\noexpand\orcidicon}}
}

\begin{document}
\title{Whispers of Supergravity in Gravitational Wave Backgrounds:\\ 
Determining the Gravitino Mass from Cosmic Thermal History}

\author{Angus Spalding \orcidB{}}
\email{angus.spalding1@gmail.com}
\affiliation{School of Physics and Astronomy, University of Southampton,
Southampton SO17 1BJ, United Kingdom}

\author{Stephen F. King \orcidA{}}
\email{S.F.King@soton.ac.uk}
\affiliation{School of Physics and Astronomy, University of Southampton,
Southampton SO17 1BJ, United Kingdom}

\begin{abstract}

Gravitino masses above the electroweak scale provide the simplest solution to the gravitino problem, but such large mass scales lie far beyond the reach of collider experiments. We show that the stochastic gravitational wave background offers a direct probe of this otherwise inaccessible regime. Despite decaying before Big-Bang Nucleosynthesis (BBN), these gravitinos naturally generate a period of early matter domination in the early universe. This non-standard epoch leaves a characteristic imprint on any primordial gravitational wave background, characterised by two frequencies corresponding to the onset and end of this phase. We demonstrate that, when early matter domination occurs while the gravitational-wave background is freely propagating, its spectral features enable the direct inference of both the gravitino mass and initial abundance, while more generally, even when the background is sourced during the early matter-dominated epoch, the gravitino mass can still be determined. Future gravitational wave observatories span a vast frequency range, enabling sensitivity to gravitino masses from the BBN bound of $\mathcal{O}(100) \text{TeV}$ up to $\mathcal{O}(10^{10}) \text{TeV}$. Already, the signal reported by NANOGrav, if primordial in origin, offers a potential probe of gravitino masses in the range $500$–$10^4 \text{TeV}$. Gravitational wave observables therefore probe an enormous region of parameter space, far beyond the reach of collider experiments. We are entering an era in which supergravity can be probed through gravitational wave backgrounds alongside collider experiments.
\end{abstract}

\maketitle

\tableofcontents

\section{Introduction}
Supersymmetry provides a well-motivated extension of the Standard Model, relating bosons and fermions and, when promoted to a local symmetry, giving rise to supergravity \cite{Golfand:1971iw, Volkov:1973ix, Salam:1974yz, Wess:1974tw, Ferrara:1976fu, Deser:1976eh, Dimopoulos:1981zb, Haber:1984rc, Nilles:1983ge, Martin:1997ns}. Beyond its field-theoretic appeal, supergravity also emerges naturally in consistent formulations of string theory and its higher-dimensional completion, M-theory \cite{Green:1984sg, Schwarz:1982jn, Candelas:1985en, Polchinski:1995mt, Witten:1995ex, Alvarez-Gaume:1983ihn}. In these frameworks, gravity is incorporated in a unified manner and predicts the existence of the gravitino, the spin-$3/2$ superpartner of the graviton \cite{Nilles:1983ge, Ellis:1983ew, Chamseddine:1982jx, Barbieri:1982eh}. The gravitino occupies a central position in the cosmology of such theories: its interactions are suppressed by the Planck scale, rendering it long-lived and highly sensitive to the thermal history of the early universe.\\
A long-standing challenge in such scenarios is the gravitino problem \cite{Weinberg82, Ellis:1983ew, Moroi:1995fs, Kawasaki:2008qe, Bolz:2000fu}: relic gravitinos produced in the early universe can decay at late times and disrupt the successful predictions of Big Bang Nucleosynthesis (BBN). The simplest resolution is to consider sufficiently heavy gravitinos, whose decay rate is enhanced by their mass, ensuring that they decay before BBN begins. This requirement translates into a lower bound on the gravitino mass of order $m_{3/2} \gtrsim \mathcal{O}(100)$ TeV. While this solution is cosmologically viable, it pushes the relevant mass scale far beyond the reach of collider experiments.\\
Even if the gravitino decays before the onset of BBN, it can still play a significant role in the expansion history of the early universe. Owing to its weak, Planck-suppressed interactions, gravitinos are generically long-lived and can be produced with an appreciable abundance, either thermally or non-thermally, for example from inflaton decay or thermal freeze-out. As a non-relativistic species, their energy density redshifts more slowly than radiation, and they can therefore come to dominate the energy budget prior to their decay. This leads to a period of early matter domination, the duration of which is set by the initial abundance, mass and decay rate of the gravitino.\\
Such modifications to the expansion history are difficult to probe with conventional cosmological observables. However, they leave a direct imprint on a stochastic gravitational-wave background, providing a unique observational handle on these otherwise inaccessible epochs \cite{Ghoshal:2026ros, Datta:2025vyu, Roshan:2026yon, Pearce_2024,BARENBOIM2016430,Assadullahi_2009,Alabidi_2013,Kohri_2018,Inomata_2019, Ferreira:2025zeu, Ferreira:2026uzi, Antusch:2024ypp, del-Corral:2026ayl}. Stochastic gravitational-wave backgrounds therefore act as cosmic archivists, preserving information from epochs far beyond the reach of standard probes \cite{Caprini_2018, Bin_truy_2012, Roshan2024, van_Remortel_2023, Christensen_2018}. In the standard cosmological history, the universe is radiation dominated after reheating, implying that gravitational-wave energy densities redshift in step with the background and thus remain approximately constant. Any deviation from radiation domination disrupts this balance, as changes in the equation of state modify the relative redshifting between gravitational waves and the dominant energy component, imprinting a characteristic distortion in the observed spectrum.\\
The majority of gravitational wave background sources are transient meaning they are no longer being sourced in our context, in particular primordial tensor modes from inflation \cite{Caprini_2016,Caprini_2018,Caldwell:2018giq,Boyle_2008,Boyle:2005se,Barman:2023ktz}, first-order phase transitions \cite{Caprini_2016, Hindmarsh:2015qta, Crawford:2024nun, Barman:2026kab}, and annihilating domain walls \cite{Vilenkin:1981zs, Vachaspati:1984gt, Blanco_Pillado_2017, Hiramatsu_2010, Barman:2022yos, Barman:2023fad} are among the most studied sources in this scenario. This is unlike for instance domain walls before annihilation \cite{Vilenkin:1981zs, Vilenkin:2000jqa} and cosmic strings \cite{Ghoshal:2025iil, Battye:2026whd, Kibble:1976sj, Vilenkin:1981zs, Vilenkin:2000jqa, Hindmarsh:1994re}. The study of stochastic gravitational-wave backgrounds has become a central focus of modern cosmology, with a broad network of current and next-generation experiments targeting frequencies spanning over twelve orders of magnitude, and with sufficient strain sensitivity to probe the physical conditions of the universe within its first second \cite{Kite:2020uix, LIGOScientific:2016aoc, LIGOScientific:2016sjg, LIGOScientific:2017bnn, LIGOScientific:2017vox, LIGOScientific:2017ycc, LIGOScientific:2017vwq, Badurina:2021rgt, Graham:2016plp, Graham:2017pmn, Badurina:2019hst, Punturo:2010zz, Hild:2010id, LIGOScientific:2016wof, Reitze:2019iox, Baker:2019nia, AEDGE:2019nxb, Sesana:2019vho, Garcia-Bellido:2021zgu, Carilli:2004nx, Janssen:2014dka, Weltman:2018zrl, EPTA:2015qep, EPTA:2015gke, NANOGRAV:2018hou, Aggarwal:2018mgp, NANOGrav:2020bcs}. The first direct detection of gravitational waves by the LIGO–Virgo collaboration \cite{LIGOScientific:2016aoc, LIGOScientific:2016sjg} established the observational era of gravitational-wave astronomy, while pulsar timing array (PTA) collaborations have recently reported compelling evidence for a stochastic background at nanohertz frequencies \cite{Carilli:2004nx, Janssen:2014dka, Weltman:2018zrl, EPTA:2015qep, EPTA:2015gke, NANOGrav:2023gor, NANOGrav:2023hvm}.\\
In this work, we demonstrate that a period of early matter domination is a generic consequence across a wide region of gravitino parameter space. We show that generic primordial gravitational wave backgrounds \footnote{In this work we consider only primordial gravitational wave sources and not astrophysical sources.} enables the direct determination of the gravitino mass. This makes it special among long-lived particles where the direct determination of the mass is not realised in cases studied so far in the literature.\\
\textit{This paper is organised as follows:} In section \ref{sec:gravitino} we review the gravitino and identify the region of parameter space in which early matter domination can occur, leading to an observable imprint in any gravitational wave background. In section \ref{sec:GWB} we analyse this imprint and show that it can be used to directly determine the gravitino mass and initial abundance in a direct mapping. We also map the parameter space to upcoming gravitational wave experiments finding the detectable range of gravitino masses and initial abundances. Finally, we conclude in section \ref{sec:conclusion}.
\section{Gravitinos in the Early Universe}
\label{sec:gravitino}
Supersymmetry provides a well-motivated extension of the Standard Model, relating bosonic and fermionic degrees of freedom \cite{Golfand:1971iw, Volkov:1973ix, Salam:1974yz, Wess:1974tw, Ferrara:1976fu, Deser:1976eh, Dimopoulos:1981zb, Haber:1984rc, Nilles:1983ge, Martin:1997ns}. When promoted to a local symmetry, it leads to supergravity, in which gravity is incorporated consistently \cite{Green:1984sg, Schwarz:1982jn, Candelas:1985en, Polchinski:1995mt, Witten:1995ex, Alvarez-Gaume:1983ihn}. A key prediction of supergravity is the existence of the gravitino, the spin-$3/2$ superpartner of the graviton \cite{Nilles:1983ge, Ellis:1983ew, Chamseddine:1982jx, Barbieri:1982eh}. The gravitino's interactions are suppressed by the Planck scale, making it generically long-lived and cosmologically relevant \cite{Dudas:2018npp, Moroi:1995fs}
\begin{equation}
    \Gamma = C \frac{m_{3/2}^3}{M_{\rm Pl}^2}
    \label{eq:decayrate}
\end{equation}
where $M_{\rm Pl}$ is the unreduced Planck mass. The dimensionless coefficient $C$ encodes the number of kinematically accessible decay channels. We consider a wide range of gravitino masses and work in the regime $m_{3/2} > m_{\rm MSSM}$, such that all MSSM decay channels are kinematically accessible. This corresponds to a high-scale supersymmetric framework in which supersymmetry is not invoked to address the electroweak hierarchy problem, but is instead motivated by its unifying structure relating matter and gauge degrees of freedom and by its role in consistent quantum theories of gravity. \footnote{We note that our results will exhibit only a mild dependence on $C$.} In the limit where all final-state masses are negligible compared to the gravitino mass, one finds\cite{Moroi:1995fs}
\begin{equation}
    C = \frac{1}{4}\left(N_V + \frac{1}{12} N_\chi \right),
\end{equation}
where $N_V$ and $N_\chi$ are the numbers of vector and chiral multiplets, respectively. For the Minimal Supersymmetric Standard Model (MSSM), $N_V = 12$ and $N_\chi = 17$, giving  $C_{\rm MSSM} \simeq 3.35$. This is the value we adopt in the analysis that follows.\\
The decay of the gravitino must occur before the onset of Big Bang Nucleosynthesis (BBN) in order not to spoil the successful predictions of light element abundances. This requirement can be imposed by demanding that the decay rate exceeds the Hubble expansion rate at $T \sim 1\,\text{MeV}$,
\begin{equation}
    \Gamma \gtrsim H(T_{\rm BBN}) \, .
\end{equation}
During radiation domination, the Hubble rate is given by
\begin{equation}
    H(T) = 1.66 \sqrt{g_*} \frac{T^2}{M_{\rm Pl}} \, ,
\end{equation}
where $g_* = 10.75$ at $T \sim 1\,\text{MeV}$. Evaluating at this temperature and using Eq.~\eqref{eq:decayrate}, one obtains
\footnote{In some models of no-scale SUGRA inflation, where SUSY is broken by a Polonyi term, 
there may also be an upper bound on the gravitino mass of 
$m_{3/2} \lesssim 100\,\text{TeV}$~\cite{Romao:2017uwa}.
Observable gravitational wave signals in ground based interferometers can also emerge from these kind of models~\cite{CrispimRomao:2023fij}. }

\begin{equation}
    m_{3/2} \gtrsim 27\,\text{TeV} \, .
\end{equation}
The full numerical calculation of this increases this bound to $m_{3/2}\gtrsim \mathcal{O}(100)\,\text{TeV} $. This bound is already beyond the reach of particle colliders, motivating the need for alternative detection mechanisms.
\subsection{A Gravitino-Dominated Era}
We consider a gravitino of mass $m_{3/2}$ that is decoupled from the thermal bath while still relativistic and subsequently evolves independently of the radiation plasma. The comoving number density of gravitinos is characterised by its initial yield $Y_i \equiv n_{3/2}/s$, where $n\ (s)$ is the number (entropy) densities respectively. We will treat the initial abundance as a free parameter assuming they are produced by a thermal or non-thermal mechanism, the most natural being inflaton decay or thermal freeze-out\cite{Kawasaki:2006gs, Takahashi:2007tz, Takahashi:2007gw, Nakayama:2012hy}. When the species behaves as matter the corresponding energy densities of matter and radiation are
\begin{equation}
    \rho_{3/2}= m_{3/2}\,n_X = m_{3/2}\,Y_i\,s,
    \qquad
    \rho_R = \frac{\pi^2}{30}\,g_*\,T^4,
\end{equation}
with $s = (2\pi^2/45)\,g_*\,T^3$ assuming no entropy dumps occur. The ratio then evolves as
\begin{equation}
    \frac{\rho_{3/2}}{\rho_R}
    = 
    \frac{m_{3/2}\,Y_i\,s}{\rho_R}
    =
    \frac{4}{3}\,Y_i\,\frac{m_{3/2}}{T},
\end{equation}
which grows as the temperature decreases. Matter domination occurs when $\rho_{3/2} = \rho_R$, defining
\begin{equation}
    T_{\rm dom} \;\simeq\; \frac{4}{3}\,Y_i\,m_{3/2}.
    \label{eq:Tdom_general_Y}
\end{equation}
When numerics were done with Boltzmann equations this was found to be the true value \cite{Ghoshal:2026ros}.\\
Having found the temperature at which the particle dominates the energy budget of the universe we now look to find when it stops dominating the universe returning to radiation domination. This calculation is much simpler as we simply match the decay rate to the Hubble parameter at the temperature after gravitino decay, $T_{\rm end}$,
\begin{equation}
    \Gamma_{3/2} = H(T_{\mathrm{end}})
    = \sqrt{\frac{8\pi^3 g_*}{90}}\,
      \frac{T_{\mathrm{end}}^2}{M_{\mathrm{Pl}}}.
\end{equation}

\begin{figure}[h!]
\centering
\includegraphics[width=1\linewidth]{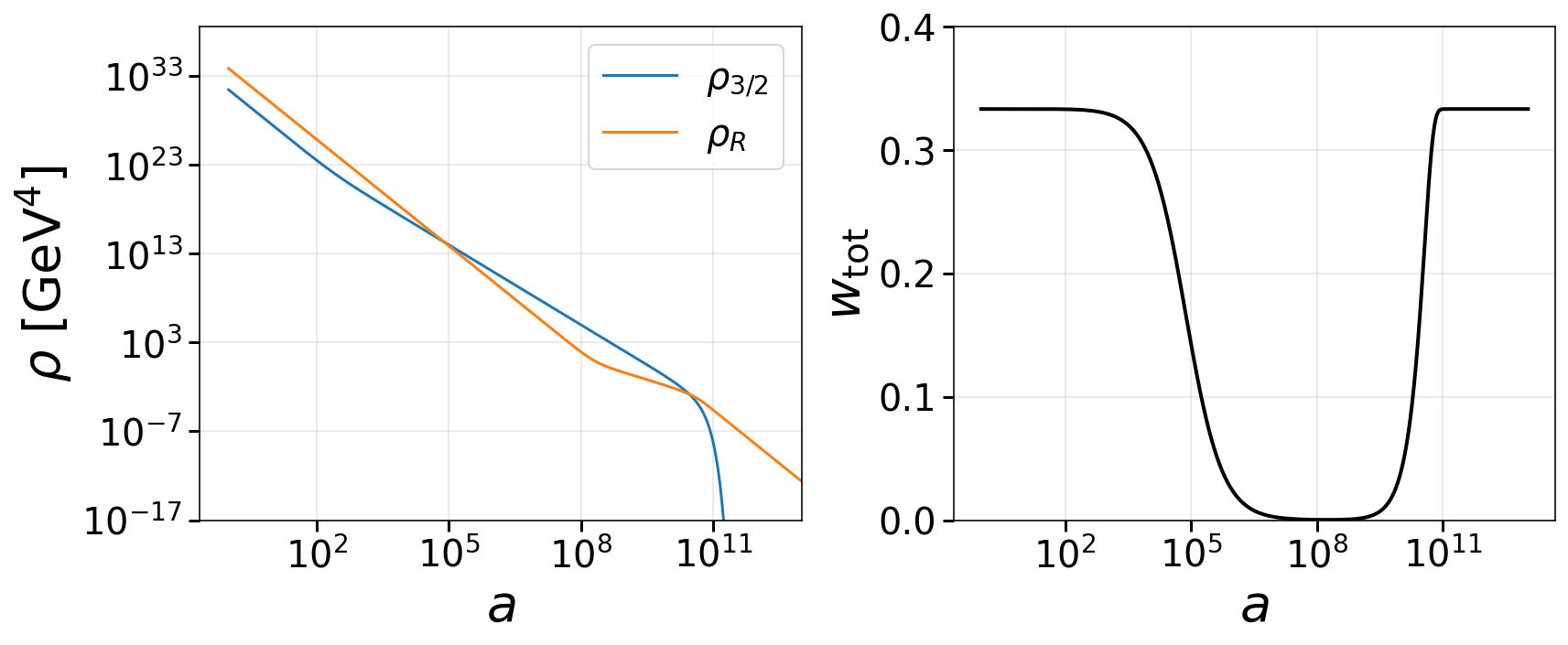} 
\caption{\it Benchmark result for early gravitino domination. \textbf{Left Panel:} shows the evolution of the energy densities of the gravitino and the MSSM thermal bath. \textbf{Right Panel:} shows the evolution of the equation of state parameter. Parameters are $m_{3/2}=10^{5}\ \text{TeV},\ Y_i=10^{-3}$. Initial temperature is taken to be $100$ times the mass of the gravitino, however the results are independent on this, provided the initial temperature is much greater than the gravitino mass. Further discussion given in the text.}
    \label{fig:Benchmark}
\end{figure}

Rearranging for $T_{\mathrm{end}}$ gives,
\begin{equation}
    T_{\mathrm{end}}
    =
    \left(\frac{90}{8\pi^3 g_*}\right)^{1/4}
    \sqrt{\Gamma_{3/2}\,M_{\mathrm{Pl}}}.
\end{equation}
When the numerics are done again the true dependence was found \cite{Ghoshal:2026ros},
\begin{equation}
    T_{\rm end} = 0.15\,\sqrt{\Gamma_{3/2} M_{\rm Pl}}
    \label{eq:Tend}
\end{equation}
The factor in front differs roughly by a factor of three to the analytical approximation but with identical scaling and dependence on parameters. To ensure that the gravitino decays only after dominating the energy density of the universe, the decay temperature must be lower than the temperature at which gravitino domination begins,
\begin{equation}
    T_{end}<T_{dom}\Rightarrow \frac{\Gamma_{3/2} M_{pl}}{Y_i^2m_{3/2}^2}<79\ .
    \label{eq:Tcondition}
\end{equation}
This expression makes explicit the dependence on both the initial abundance $Y_i$ and the particle mass $m_{3/2}$, and is applicable to both thermal and non-thermal production scenarios. This would normally require unnaturally small couplings however for the gravitino this is very naturally achieved as the gravitino only decays gravitationally. Applying the gravitino decay rate $\Gamma\simeq m_{3/2}^3/M_{pl}^2$ one can see the dependence on this condition being, 
\begin{equation}
    \frac{\Gamma_{3/2} M_{pl}}{Y_i^2m_{3/2}^2}\lesssim 79\Rightarrow \frac{m_{3/2}}{Y_i^2M_{pl}}\lesssim79
\end{equation}
and one can immediately see how natural this is for gravitinos whose mass is expected to be well below the Planck scale. The exact condition on the initial abundance can now be written as
\begin{equation}
    Y_i \gtrsim 10^{-8} \sqrt{\frac{m_{3/2}}{100\,\text{TeV}}} \, .
\end{equation}
For representative values,
\begin{equation}
    m_{3/2} = 100\,\text{TeV} \;\Rightarrow\; Y_i \gtrsim 10^{-8}
    \quad,\quad
    m_{3/2} = 10^{9}\,\text{TeV} \;\Rightarrow\; Y_i \gtrsim 3.2 \times 10^{-5} \, .
\end{equation}
These values of the initial abundance are well motivated in scenarios where gravitinos are produced non-thermally, for example via inflaton decay. To find this period numerically one must solve the Boltzmann equations for the process, which are written below.
\begin{equation}
    \dot{\rho}_{3/2}+3H(w(t)+1)\rho_{3/2}=-\Gamma_{3/2}\rho_{3/2}
\end{equation}
\begin{equation}
    \dot{\rho_{R}}+4H\rho_{R}=\Gamma_{3/2}\rho_{3/2}
\end{equation}
where $w(t)$ denotes the equation of state parameter for the gravitino that interpolates between $w=1/3$ when $T\gg m_{3/2}$ and $w=0$ when $T\ll m_{3/2}$. We change variables to the scale factor $a$ and provide a typical benchmark of the evolution of these variables in Fig \ref{fig:Benchmark}.\\
At early times, when $T \gg m_{3/2}$, the gravitino is relativistic and its energy density redshifts as radiation, corresponding to an equation of state of the universe $w_{\rm tot} = 1/3$. As the temperature drops below the gravitino mass, $T \lesssim m_{3/2}$, the species becomes non-relativistic and its energy density scales as $a^{-3}$, while radiation continues to scale as $a^{-4}$. As a result, the gravitino can come to dominate the total energy density and the equation of state parameter of the universe is $w_{\rm tot}=0$. Once the decay rate becomes comparable to the Hubble expansion rate, $\Gamma \sim H$, the gravitino decays and its energy is transferred back into radiation, restoring radiation domination. The equation of state therefore evolves from $w_{\rm tot} = 1/3$ in the initial radiation-dominated phase to $w_{\rm tot} = 0$ during the period of gravitino domination, before returning to $w_{\rm tot} = 1/3$ after decay and crucially before BBN.
\subsection{Gravitino Production Mechanisms}

So far, we have treated the initial gravitino abundance as a free
parameter. Here we briefly discuss a few production mechanisms
and determine when they can generate an abundance satisfying
\begin{equation}
    Y_i \gtrsim 10^{-8}
    \sqrt{\frac{m_{3/2}}{100\,\text{TeV}}}\, .
    \label{eq:required-gravitino-yield}
\end{equation}
We assume throughout that there is no significant entropy production
after gravitino production until the gravitinos themselves decay.

\begin{itemize}
    \item \textbf{Thermal freeze-in.}

    Gravitinos are produced through scatterings of particles in the
    thermal plasma after reheating. For a reheating temperature well
    above the relevant particle masses and large gravitino masses that we consider here, the resulting yield is \cite{Kohri:2005wn, Kawasaki:2004qu}
    \begin{equation} Y_{3/2}^{\rm th} \simeq 1.9 \times 10^{-12} \left(\frac{T_R}{10^{10}\,\mathrm{GeV}}\right) \left[ 1 + 0.045 \ln\!\left(\frac{T_R}{10^{10}\,\mathrm{GeV}}\right) \right] \left[ 1 - 0.028 \ln\!\left(\frac{T_R}{10^{10}\,\mathrm{GeV}}\right) \right], \label{eq:thermal-gravitino-yield} \end{equation}
    therefore to reach our bound for initial abundance of $Y_{3/2}^{\rm min}=10^{-8}$
    requires a reheating temperature of 
    \begin{equation} T_R^{\rm min} \simeq 5 \times 10^{13}\,\mathrm{GeV}. \label{eq:minimum-reheating-temperature} \end{equation}
    Thus, thermal production can generate the required abundance, but
    generally requires a rather high reheating temperature. Taking the maximum reheating temperature allowed $T_{RH}\lesssim 10^{16}$ GeV \cite{Gong:2015qha}  this puts the bound on gravitino mass than can come to dominate and produce $m_{3/2}^{\rm max}\simeq 3.6\times 10^6$ TeV. Gravitinos above this mass must be produced non-thermally to dominate the universe before decaying.

    \item \textbf{Inflaton decay.}

    The most natural production mechanism is that of inflaton decay. During inflation the energy density is dominated by the inflaton which then decays into radiation particles reheating the universe. This can then produce a gravitino yield
    \begin{equation}
        Y_{3/2}^{\phi}
        \simeq
        \frac{3}{4}\,
        N_{3/2}\,\text{Br}_{3/2}\,
        \frac{T_R}{m_\phi},
        \label{eq:inflaton-gravitino-yield}
    \end{equation}
    where $m_\phi$ is the inflaton mass, $\text{Br}_{3/2}$ is the
    branching fraction into a final state containing gravitinos, and
    $N_{3/2}$ is the number of gravitinos produced per decay. The
    required abundance is obtained from eEq \ref{eq:required-gravitino-yield} when
    \begin{equation}
        N_{3/2}\,\text{Br}_{3/2}
        \gtrsim
        1.33 \times 10^{-8}
        \frac{m_\phi}{T_R}
        \sqrt{\frac{m_{3/2}}{100\,\text{TeV}}}\, .
        \label{eq:inflaton-production-condition}
    \end{equation}
    Numerically, this becomes
    \begin{equation}
        N_{3/2}\,\text{Br}_{3/2}
        \gtrsim
        1.33 \times 10^{-5}
        \left(\frac{m_\phi}{10^{13}\,\text{GeV}}\right)
        \left(\frac{10^{10}\,\text{GeV}}{T_R}\right)
        \sqrt{\frac{m_{3/2}}{100\,\text{TeV}}}\, .
    \end{equation}
    For the direct pair-production channel
    $\phi\rightarrow2\psi_{3/2}$, one has $N_{3/2}=2$ and the decay is
    kinematically allowed only for $m_\phi>2m_{3/2}$. 
\end{itemize}

Other possible sources include production during preheating,
gravitational particle production at the end of inflation, and the
late decay of heavier hidden-sector states such as moduli. 
\section{Gravitational Wave Background Signatures}
\label{sec:GWB}
In this section we develop the framework for probing the gravitino using gravitational wave backgrounds and apply it to the gravitino. In previous analyses \cite{Datta:2025vyu, Ghoshal:2026ros}, the mapping involved the initial abundance, mass, and decay rate as inputs to two gravitational-wave observables. By considering the gravitino as the dominant particle, this analysis is refined, since the parameter space reduces to a direct mapping between the initial abundance and the gravitino mass. An additional advantage is that, since the gravitino decays only through gravitational interactions, the condition for early matter domination in Eq.~\ref{eq:Tcondition} is naturally satisfied.
\subsection{Gravitational Wave Signatures of Early Matter Domination}
\begin{figure}[h!]
\centering
\includegraphics[width=1\linewidth]{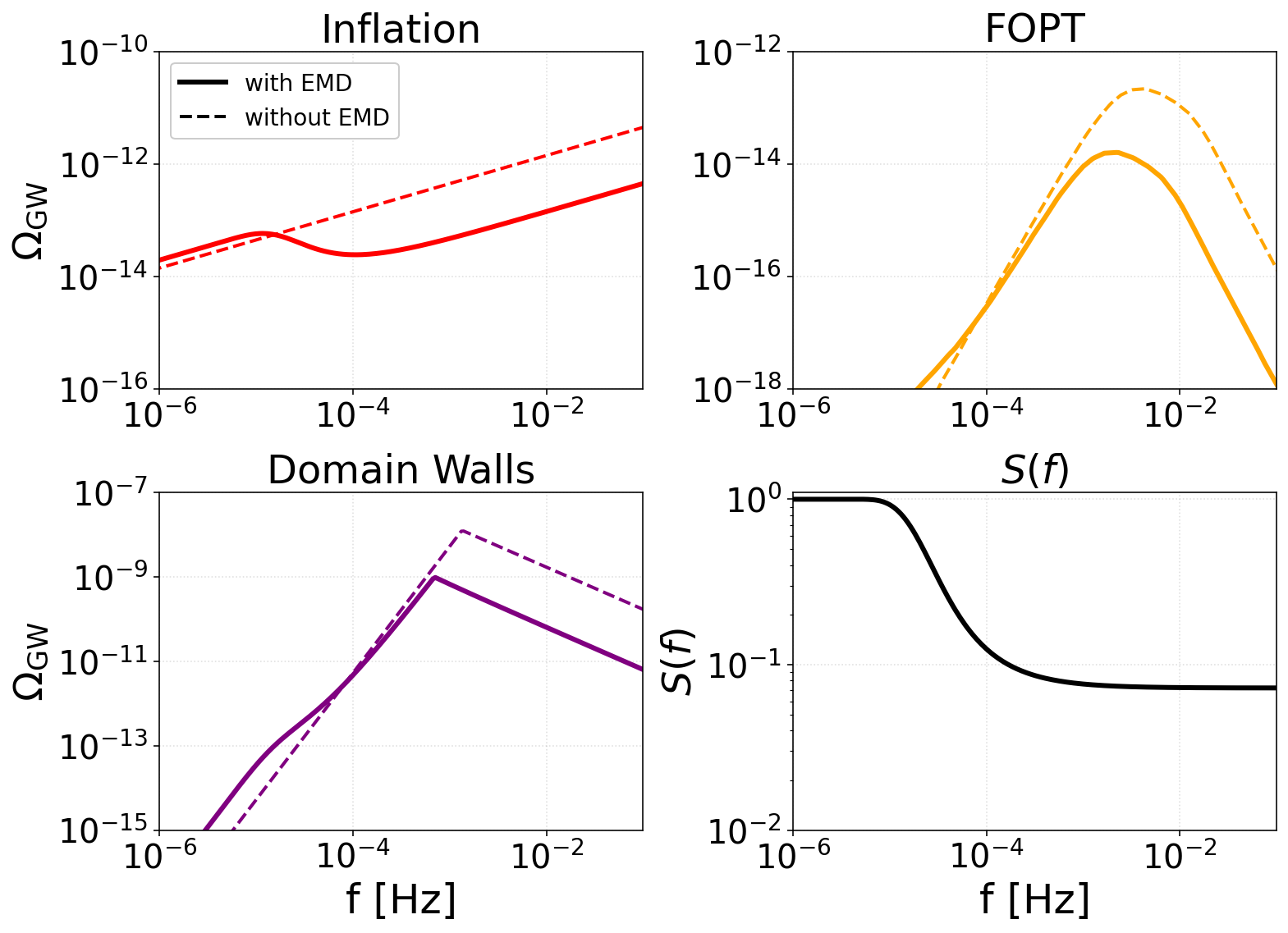} 
\caption{\it Benchmark GWBs with (solid lines) and without (dashed lines) an early matter-dominated era. Results are shown for inflationary source (red), first-order phase transitions source (orange), and domain wall source (purple). Parameters for each benchmark shown in table \ref{tab:bench}. In each case a clear features can be seen and the frequencies corresponding to EMD can be seen. The fourth panel shows the function S defined in Eq~\ref{eq:S}. Further discussion is given in the text.}
    \label{fig:bench1}
\end{figure}

The quantity of interest is the gravitational wave background spectrum, defined as the fractional energy density in gravitational waves per logarithmic frequency interval,
\begin{equation}
    \Omega_{\rm GW}(a,f)
    \equiv
    \frac{1}{\rho_{\rm tot}(a)}
    \frac{d\rho_{\rm GW}(a)}{d\ln f}.
\end{equation}
The goal will be to figure out how this quantity changes with equation of state parameter. We give this quantity dependence on not just the frequency but the scale factor as this will be our evolution parameter. We assume that there is a gravitational wave background that is no longer being sourced and is freely propagating so that the only change is the cosmological evolution due to the change of the equation of state of the universe.\\
Each observed frequency corresponds to a particular gravitational-wave mode. This mode does not immediately behave as a radiation component after production. While its physical wavelength is larger than the Hubble radius (i.e. the mode is outside the horizon), the tensor perturbation is effectively frozen and does not propagate as a wave. In this regime the usual radiation-like redshifting description is not
applicable. As the universe expands, the Hubble radius grows relative to the wavelength, and the mode eventually enters the horizon when its physical scale becomes comparable to the Hubble scale. From this point onward the mode oscillates and propagates freely, redshifting like radiation. Therefore, the evolution of the gravitational-wave background begins only after horizon entry, and $\Omega_{GW}$ evolves only once the corresponding mode is inside the horizon.\\
Horizon entry occurs when the physical wavelength of the mode becomes comparable to the Hubble radius. Equivalently, this condition can be written by comparing the physical wavenumber $k/a$ with the expansion rate $H(a)$. The transition from superhorizon to subhorizon behaviour is defined by
\begin{equation}
    k=a_{ent}H(a_{ent}),\quad k=2\pi f_0a_0
\end{equation}
where $f_0$ is the frequency as measured today and $a_0$ is the scale factor measured today which we calculate from our Boltzmann code assuming that between the end of the Boltzmann evolution and today there are no non-negligible entropy dumps,
\begin{equation}
    a_0=a_f\frac{T_f}{T_0}\left(\frac{g_*(T_f)}{g_*(T_0)}\right)^{\frac{1}{3}},
    \label{eq:Tf}
\end{equation}
where $a_f$ is the final scale factor of the evolution of Boltzmann equations, $T_f$ and $T_0$ are the final temperature of Boltzmann's and the temperature today respectively, $g_*$ is the degrees of freedom for the corresponding times.\\
This equation implicitly defines the entry scale factor for each frequency. In other words, the function $a_{ent}(f)$ provides the mapping between frequency space and cosmic time: higher-frequency modes satisfy this condition at earlier times (smaller scale factor), while lower-frequency modes enter later.\\
Once a mode is inside the horizon, it behaves as a freely propagating gravitational wave. The fluid equations for each of the species is, 
\begin{equation}
    \dot{\rho}_{GW}+4H\rho_{GW}=0,\quad \dot{\rho}_{tot}+3H(w(a)+1)\rho_{tot}=0
\end{equation}
so differentiating with e-folds gives,
\begin{equation}
    \frac{d\ ln\ \rho_{tot}}{d\ ln\ a}=-3(w(a)+1)\quad \frac{d\ ln\ \rho_{GW}}{d\ ln\ a}=-4
\end{equation}
Meaning the gravitational wave background $\Omega_{GW}$ changes as,
\begin{equation}
    \frac{d\ ln\ \Omega_{GW}}{d\ ln\ a}=3w(a)-1
\end{equation}
We now integrate this equation to obtain the full evolution of the spectrum. Evolution starts at the horizon-entry scale factor $a_{ent}(f)$, we therefore integrate from this point to some later time after we return to radiation domination. Integrating both sides and exponentiation gives,
\begin{equation}
    \Omega_{GW}(a_f,f)= \Omega_{GW}^{RD}(f)Exp\left[\int_{ln\ a_{ent}(f)}^{ln\ a_f}(3w(a)-1)\, d\ln a\right]
\end{equation}
During radiation domination ($w = 1/3$), the gravitational wave background remains constant. As the universe undergoes a period of early matter domination ($w < 1/3$), the gravitational wave energy density is suppressed relative to the background, producing a characteristic drop in the spectrum. Once the universe returns to radiation domination, the spectrum again becomes constant until the onset of the standard late-time matter-dominated era. The treatment above approximates horizon entry as an instantaneous transition to radiation-like redshifting. To account
for the evolution through this transition, we consider the tensor
mode equation,
\begin{equation}
    h_k''+2\mathcal{H}h_k'+k^2h_k=0,
    \qquad \mathcal{H}=aH,
\end{equation}
where primes denote conformal-time derivatives, the source has
switched off, and tensor anisotropic stress is neglected. The
non-decaying solution is constant for $k\ll aH$, while for
$k\gg aH$ it oscillates with an envelope proportional to $a^{-1}$,
recovering the energy-density scaling used above. The evolution
between these limits fixes the subhorizon normalisation. For an
approximately constant equation of state during horizon entry,
this gives a correction to the instantaneous-entry prescription
\cite{Boyle_2008},
\begin{equation}
C(w[k=aH])
=
\frac{\Gamma^2\!\left(\alpha+\tfrac{1}{2}\right)}{\pi}
\left(\frac{2}{\alpha}\right)^{2\alpha},\quad \alpha=\frac{2}{1+3w}
\end{equation}
where $\alpha$ should be evaluated at horizon re-entry, $k=aH$ assuming that $w$ is roughly constant during the crossing. This factor is therefore 1 for radiation domination and matter domination is $9/16$ so is an $\mathcal{O}(1)$ correction to the amplitude. We must also take into account that, in Eq.~\ref{eq:Tf}, the final temperature is modified in the presence of an early matter-dominated era compared to the purely radiation-dominated case. This arises from the entropy injection associated with the decay of the dominating species. Consequently, the frequencies at the original source are shifted by a factor depending on the initial and final comoving entropies, $S_i$ and $S_f$, respectively. The final form of the equation is therefore \cite{Ghoshal:2026ros},
\begin{equation}
    \Omega_{GW}(a_f,f)= \Omega_{GW}^{RD}(f')C(w_{ent})Exp\left[\int_{ln\ a_{ent}(f)}^{ln\ a_f}(3w(a)-1)\, d\ln a \right],\quad f'=f \left(\frac{S_f}{S_i}\right)^{1/3}\ .
    \label{eq:MDeffect}
\end{equation}
The characteristic imprint of an early matter-dominated (EMD) era on the gravitational-wave spectrum is therefore twofold: first, a leftward shift of the spectrum arising from the entropy injection at the end of the EMD period, and second, a suppression over a range of frequencies, producing a kink-like feature in the spectrum from the change in redshifting. The resulting spectrum is characterised by two frequencies corresponding to the onset and end of the EMD era \cite{Datta:2025vyu, Ghoshal:2026ros}. These frequencies constitute the relevant gravitational-wave observables, encoding the beginning and end of the matter-dominated phase. Importantly, this effect is independent of the origin of the gravitational-wave background, so long as the source is transient and is not being produced at the time of the EMD period. In each case, the break and domination frequencies are determined by the underlying Lagrangian parameters and the initial abundance of the dominating species. 
A useful quantity for identifying the frequencies at which deviations from radiation domination occur is defined as
\begin{equation}
    S(f)=C(w_{\mathrm{ent}})\exp\left[\int_{\ln a_{\mathrm{ent}}(f)}^{\ln a_f}(3w(a)-1)\, d\ln a \right]\ .\label{eq:S}
\end{equation}
This expression shows that modes entering the horizon after the period of early matter domination experience no suppression, since \(S=1\) in this regime. In contrast, modes that enter during or before the early matter-dominated era are suppressed, with modes spending a longer time in the early matter phase undergoing greater dilution.\\
To demonstrate the broad applicability of this framework, we consider several benchmark gravitational wave backgrounds generated during radiation domination, followed by a period of early matter domination driven by gravitino domination. In each scenario, a distinct signature emerges through deviations from pure radiation domination. The resulting spectra are shown in Fig.~\ref{fig:bench1}, while the parameters used to generate the benchmarks are listed in Table~\ref{tab:bench}. We also include a fourth panel displaying the function \(S(f)\), which captures both the deviation from radiation domination and the associated suppression.
\begin{table}[h!]
\centering
\footnotesize
\setlength{\tabcolsep}{1.5pt}
\renewcommand{\arraystretch}{0.80}

\begin{tabular}{|l l|}
\hline
\textbf{Source} & \textbf{Parameters} \\
\hline

Inflation 
& $\epsilon_{\rm inf} = 5\times 10^{-5},\; n_T = 0.5$ \\

FOPT 
& $\alpha = 0.2$, $\beta/H=50$, $T_n=200$ GeV, $v_w=1$ \\

Domain walls 
& $\sigma_t = 2\times10^{17} \rm GeV,\; V_t = 6\times10^{28} GeV,\; \epsilon = 0.7,\; A = 0.8$\\
\hline
\hline

$\psi_{3/2}$ & $m_{3/2}=3\times 10^5$ TeV,\ $Y_i=10^{-5}$\\
\hline
\end{tabular}

\caption{\it Benchmark parameters used for each gravitational-wave source shown in Fig.~\ref{fig:bench}. Details of the parameter choices for each benchmark scenario can be found in \cite{Ghoshal:2026ros} and in standard reviews of stochastic gravitational-wave backgrounds \cite{Roshan2024}. We intentionally do not focus on the specific gravitational wave source parameters, since the resulting signatures and the information that can be extracted are independent of the details of the underlying gravitational-wave source. The location of the early matter domination features are solely determined by the $\psi_{3/2}$ parameters.
}
\label{tab:bench}
\end{table}
Importantly, the function \(S(f)\) is determined exclusively by the period of gravitino domination. Consequently, the imprint on the gravitational wave background is independent of the production mechanism and is instead characterised solely by two frequencies corresponding to the onset and end of the early matter-dominated era. From all four panels, we identify the end of early matter domination at approximately \(f_1 \simeq 10^{-5}\,\mathrm{Hz}\), while the onset occurs around \(f_2 \simeq 10^{-3.5}\,\mathrm{Hz}\) for the benchmark considered. We now relate these observable frequencies to the underlying gravitino parameters.
\subsection{Mapping Observables to Gravitino Parameters}
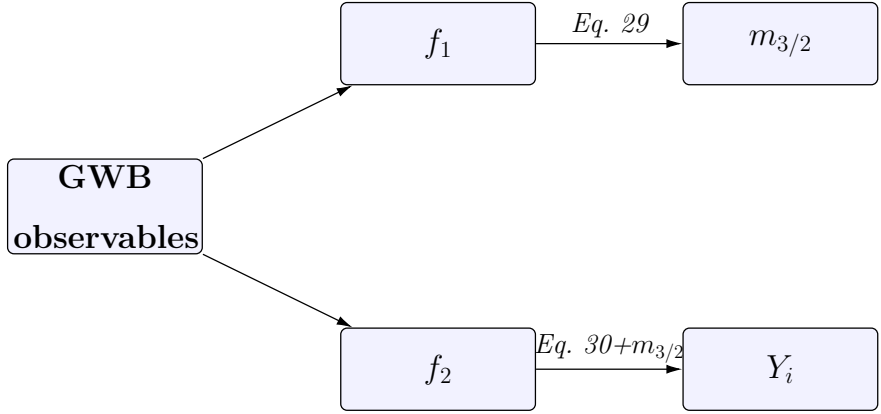
\begin{figure}[t]
    \centering
    \resizebox{0.7\linewidth}{!}{%
    \begin{tikzpicture}[
        node distance=3.0cm and 2.6cm,
        box/.style={
            rectangle, rounded corners, draw=black, fill=blue!5, thick,
            text centered, text width=4.0cm, minimum height=1.8cm,
            font=\bfseries\LARGE    
        },
        arrow/.style={thick, -{Latex[length=3.5mm,width=2mm]}},
        eqnlabel/.style={midway, above, sloped, font=\Large\itshape},
        eqnlabelV/.style={pos=0.5, left, font=\Large\itshape}
    ]

    \node[box] (gwb) {GWB\\ observables};

    \node[box, above right=1.6cm and 3.cm of gwb] (fbrk) {$f_1$};
    \node[box, below right=1.6cm and 3.cm of gwb] (fdom) {$f_2$};

    \node[box, right=3.2cm of fdom] (tdom) {$Y_i$};
    \node[box, right=3.2cm of fbrk] (tend) {$m_{3/2}$};

    \draw[arrow] (gwb) -- (fbrk);
    \draw[arrow] (gwb) -- (fdom);

    \draw[arrow] (fdom) -- node[eqnlabel] {Eq.~\ref{eq:f2}+$m_{3/2}$} (tdom);
    \draw[arrow] (fbrk) -- node[eqnlabel] {Eq.~\ref{eq:f1}} (tend);

    \end{tikzpicture}%
    }

    \caption{
        \it Schematic illustrating the connection between a gravitational-wave background (GWB) and gravitino parameters. A stochastic GWB can contain two characteristic frequencies: $f_2$, marking the onset of an early matter-dominated era, and $f_1$, corresponding to the transition back to radiation domination. From $f_1$, the gravitino mass can be inferred via Eq.~\ref{eq:f1}. Meanwhile, $f_2$ determines the combination of the gravitino mass and initial abundance through Eq.~\ref{eq:f2}. Combining this with the value of $m_{3/2}$ extracted from $f_1$ allows the initial abundance to be determined directly, thereby establishing a direct correspondence between gravitational-wave observables and the gravitino parameters.
  }
    \label{fig:GWB_inference_chain}
\end{figure}

\begin{figure}[h!]
\centering
\includegraphics[width=1\linewidth]{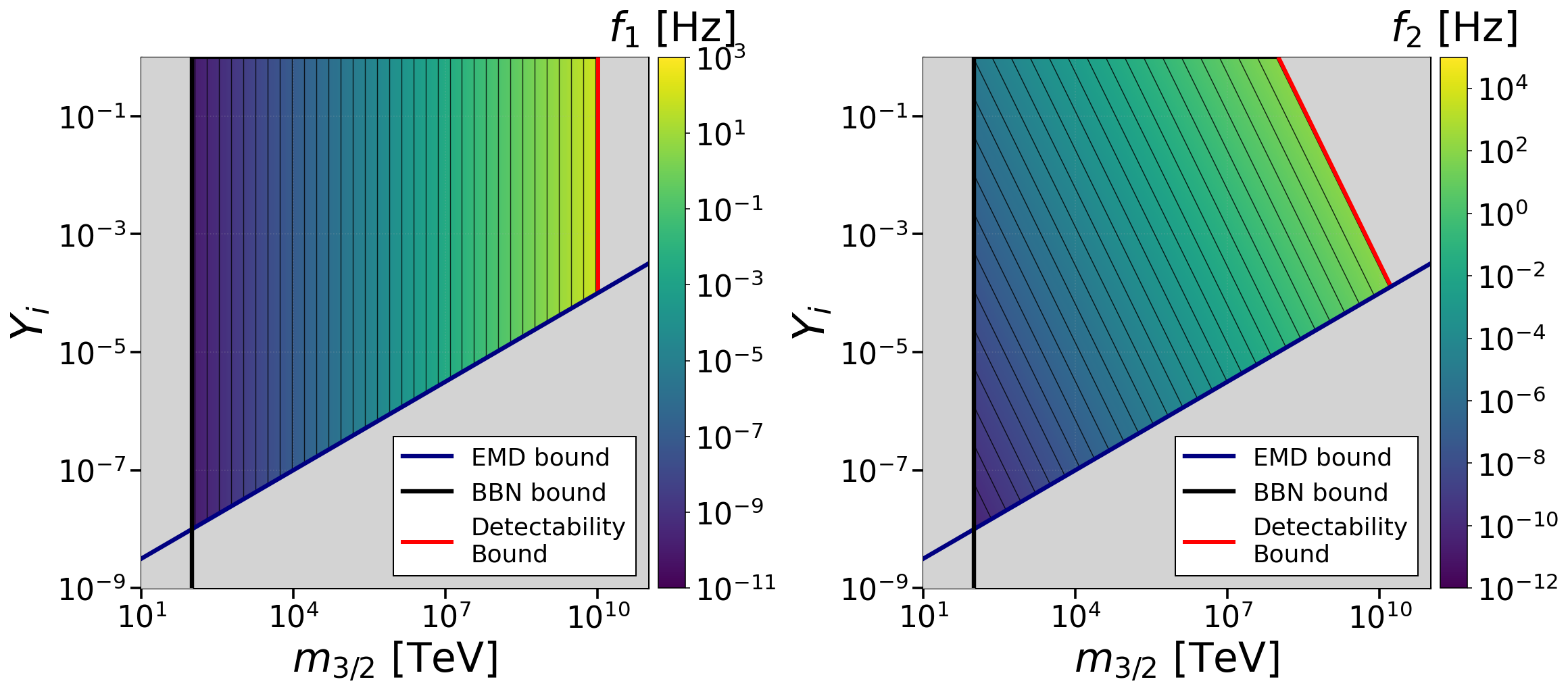} 
\caption{\it The observable frequencies characterising the spectral features of the gravitational wave background as functions of the initial abundance $Y_i$ and gravitino mass $m_{3/2}$. \textbf{Left Panel:} the frequency $f_1$, corresponding to the end of the early matter-dominated era. \textbf{Right Panel:} the frequency $f_2$, corresponding to the onset of the early matter-dominated era. In both panels, the region to the left of the red contour corresponds to frequencies accessible to upcoming gravitational-wave experiments. The region to the left of the black vertical line is excluded by Big Bang Nucleosynthesis (BBN) constraints, while the region below the cyan curve does not lead to an early matter-dominated phase and is therefore excluded from the analysis.}
    \label{fig:Fscan}
\end{figure}

\begin{figure}[h!]
\centering
\includegraphics[width=0.7\linewidth]{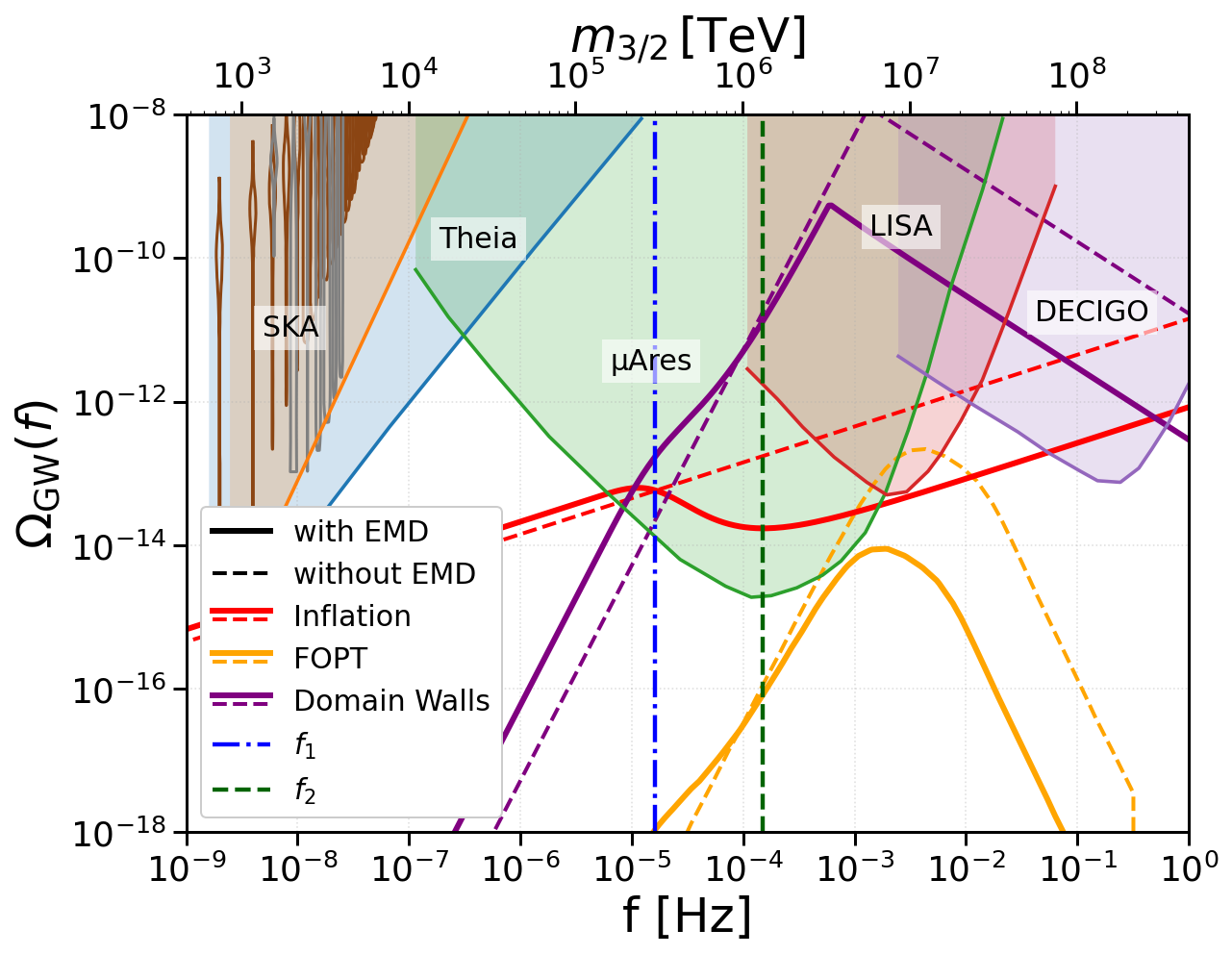} 
\caption{\it Benchmark GWBs with (solid lines) and without (dashed lines) an early matter-dominated era. Results are shown for inflationary source (red), first-order phase transitions source (orange), and domain wall source (purple). The shaded regions indicate the projected sensitivity of each detector, with the corresponding experiments labelled. Parameters for each benchmark shown in table \ref{tab:bench}. The upper x-axis indicates the location of the first feature $f_1$ for a given gravitino mass. For example we can clearly see this deviation from radiation domination corresponds to $m_{3/2}=3\times 10^{5}\ \rm TeV$. Further discussion is given in the text.}
    \label{fig:bench}
\end{figure}

To extract the characteristic frequencies quantitatively, we follow
the numerical procedure of Ref.~\cite{Ghoshal:2026ros}. The Boltzmann evolution determines
$w(a)$ and $H(a)$, from which the horizon-entry scale factor and the
suppression function $S(f)$ are evaluated for each frequency. The resulting
function is fitted using
\begin{equation}
    S_{\mathrm{fit}}(f)
    =
    \frac{1+(f/f_2)^2}{1+(f/f_1)^2},
    \qquad f_1<f_2.
    \label{eq:S_fit}
\end{equation}
This form captures the unsuppressed low-frequency plateau, the
$S(f)\propto f^{-2}$ behaviour for modes entering during matter
domination, and the suppressed high-frequency plateau:
\begin{equation}
    S_{\mathrm{fit}}(f)\simeq
    \begin{cases}
        1, & f\ll f_1,\\
        (f_1/f)^2, & f_1\ll f\ll f_2,\\
        (f_1/f_2)^2, & f\gg f_2.
    \end{cases}
\end{equation}
The intermediate scaling follows from $H\propto a^{-3/2}$ and
$k=a_{\mathrm{ent}}H(a_{\mathrm{ent}})$, giving
$a_{\mathrm{ent}}\propto f^{-2}$. The fitted frequencies characterise
the smooth transitions associated with the end and onset of early
matter domination, respectively, rather than being imposed as
instantaneous horizon scales. Repeating this procedure over the
particle parameter space gives the numerical relations between
$f_1$, $f_2$ and the underlying parameters.
The frequency $f_1$ depends only on the decay rate, as it corresponds to the end of the early matter dominated era and is therefore determined by the decay temperature, see Eq.~\ref{eq:Tend}. In contrast, $f_2$ depends strongly on the combination $Y_i m_{3/2}$, which sets the onset of the early matter dominated phase. However, when the frequency is redshifted to the present day, there is also a mild dependence on the duration of the matter dominated era, leading to a subdominant dependence on the decay rate as well. The scan for $f_1$ yields the best-fit relation \cite{Ghoshal:2026ros}
\begin{equation}
    f_1 = 20.6\,\left(\frac{\Gamma}{\rm GeV}\right)^{1/2}\,\text{Hz}
    =
    9.77\times10^{-11}
    \left(\frac{m_{3/2}}{100\,\text{TeV}}\right)^{3/2}
    \text{Hz},
    \label{eq:f1}
\end{equation}
while the scan for $f_2$ gives the best-fit relation \cite{Ghoshal:2026ros}
\begin{equation}
    f_2 = 2.77\times10^{-5}\,
    \left(\frac{Y_i m_{3/2}}{\rm GeV}\right)^{2/3}
    \left(\frac{\Gamma}{\rm GeV}\right)^{1/6}
    \text{Hz}
    =
    1.00\times10^{-5}\,
    Y_i^{2/3}
    \left(\frac{m_{3/2}}{100\,\text{TeV}}\right)^{7/6}
    \text{Hz}.
    \label{eq:f2}
\end{equation}
where in the last step we implemented the gravitino decay rate \footnote{The specifics of the best fit analysis is taken from \cite{Ghoshal:2026ros}.}. These relations provide a one-to-one mapping between the observable frequencies of the gravitational-wave background and the decay rate and product of the mass and initial abundance. We show this as a schematic in Fig \ref{fig:GWB_inference_chain}.
Scans over these characteristic frequencies, together with the bounds arising from the requirement of an early matter-dominated era and consistency with Big Bang Nucleosynthesis (BBN), as well as the experimentally accessible frequency ranges, are shown in Fig.~\ref{fig:Fscan}. This demonstrates the significant region of parameter space that will be accessible to upcoming gravitational wave experiments.
\\
\noindent Among these observables, $f_1$ is the most powerful, as it alone can be used to directly determine the gravitino mass. Inverting Eq.~\eqref{eq:f1}, the gravitino mass can be expressed in terms of the observed frequency as
\begin{equation}
    m_{3/2} = 4.71 \times 10^{6} \, f_1^{2/3} \,(100\,\text{TeV}) \, .
\end{equation}
To illustrate this, we present the benchmark spectra of Fig.~\ref{fig:bench1} together with the projected sensitivities of future experiments in Fig.~\ref{fig:bench}. Crucially, we also include an upper \(x\)-axis in Fig.~\ref{fig:bench} showing the corresponding gravitino mass, allowing the mass to be read directly from the characteristic frequency \(f_1\), which marks the departure from radiation domination.
\noindent For the domain wall and inflationary gravitational-wave benchmarks, the spectral features induced by the early matter dominated era are detectable and can be used to extract information about the gravitino. We also demonstrate a complementary scenario for gravitational waves sourced by a first-order phase transition, where the suppression caused by the early matter dominated period can render the gravitational-wave background undetectable. Consequently, an early matter dominated epoch can reopen regions of parameter space that would otherwise be considered excluded. Importantly, as can be seen from this figure, the kink occurs at the same location independent of the source. \\
This has an implication for detection: if the spectral break associated with the end of the early matter dominated era is observed in a primordial gravitational wave background, then $f_1$ is automatically accessible, allowing the gravitino mass to be directly inferred. Conversely, if no deviation from radiation domination is observed, the entire region of parameter space satisfying Eq.~\ref{eq:Tcondition} within the frequency range probed by a given experiment can be excluded.\\
Using the characteristic frequency ranges relevant for future gravitational-wave observatories, their sensitivity can be mapped directly onto the corresponding gravitino mass scale. The resulting ranges are summarised in Table~\ref{tab:gwb_mass}.
\begin{table}[h]
\centering
\begin{tabular}{c | c | c}
\hline
Experiment & $f$ [Hz] & $m_{3/2}$ [TeV] \\
\hline
Pixie & $<10^{-9}$ & Down to BBN Bound  \\
SKA \& NANOGrav     & $10^{-9}$ -- $10^{-7}$   & $4.7\times10^{2}$ -- $1.0\times10^{4}$ \\
Theia   & $10^{-9}$ -- $10^{-5}$   & $4.7\times10^{2}$ -- $2.2\times10^{5}$ \\
$\mu$Ares & $10^{-7}$ -- $10^{-2}$ & $1.0\times10^{4}$ -- $2.2\times10^{7}$ \\
LISA    & $10^{-4}$ -- $10^{-1}$   & $1.0\times10^{6}$ -- $1.0\times10^{8}$ \\
DECIGO  & $10^{-2}$ -- $10$        & $2.2\times10^{7}$ -- $2.2\times10^{9}$ \\
ET      & $1$ -- $10^{2}$          & $4.7\times10^{8}$ -- $1.0\times10^{10}$ \\
CE      & $5$ -- $10^{2}$          & $1.0\times10^{9}$ -- $1.0\times10^{10}$ \\
\hline
\end{tabular}
\caption{\it Gravitino mass ranges probed by current and future gravitational-wave observatories, obtained by mapping their characteristic frequency sensitivities onto the decay-induced gravitational-wave signal. The broad frequency coverage of these experiments translates into sensitivity to gravitino masses spanning more than eight orders of magnitude, ranging from $\mathcal{O}(10^2)\,\mathrm{TeV}$ for PIXIE and PTA experiments, up to $\mathcal{O}(10^{10})\,\mathrm{TeV}$ for ground-based interferometers such as Einstein telescope and cosmic explorer.}
\label{tab:gwb_mass}
\end{table}
This demonstrates that different experiments probe widely separated regions of parameter space, with PTA experiments sensitive to comparatively light gravitinos, while ground-based interferometers such as ET probe extremely heavy masses. \\
An important point to emphasise is that if this feature were observed, it would not uniquely identify the gravitino as its source. More generally, such a signal would indicate the presence of an early matter-dominated era and if sourced by a long-lived particle would determine the corresponding decay rate and a fixed combination of mass and initial abundance. However, the converse statement is considerably stronger. If a primordial GWB is detected in the next generation of experiments and this feature is absent, then the parameter space shown in Fig.~\ref{fig:Fscan} would be ruled out, subject to astrophysical backgrounds. This exclusion spans several orders of magnitude in both gravitino mass and initial abundance, making even a null result highly informative. 
More broadly, gravitational wave backgrounds provide, to the authors' knowledge, the only observational probe of high-scale supergravity. Even a null result can place significant constraints on the allowed gravitino parameter space. We are entering an era in which gravitational wave observatories are beginning to probe high-scale physics beyond the Standard Model in regimes far beyond the reach of collider experiments.
\subsection{A Benchmark Fit to NANOGrav Data}
\begin{figure}[h!]
\centering
\includegraphics[width=0.49\linewidth]{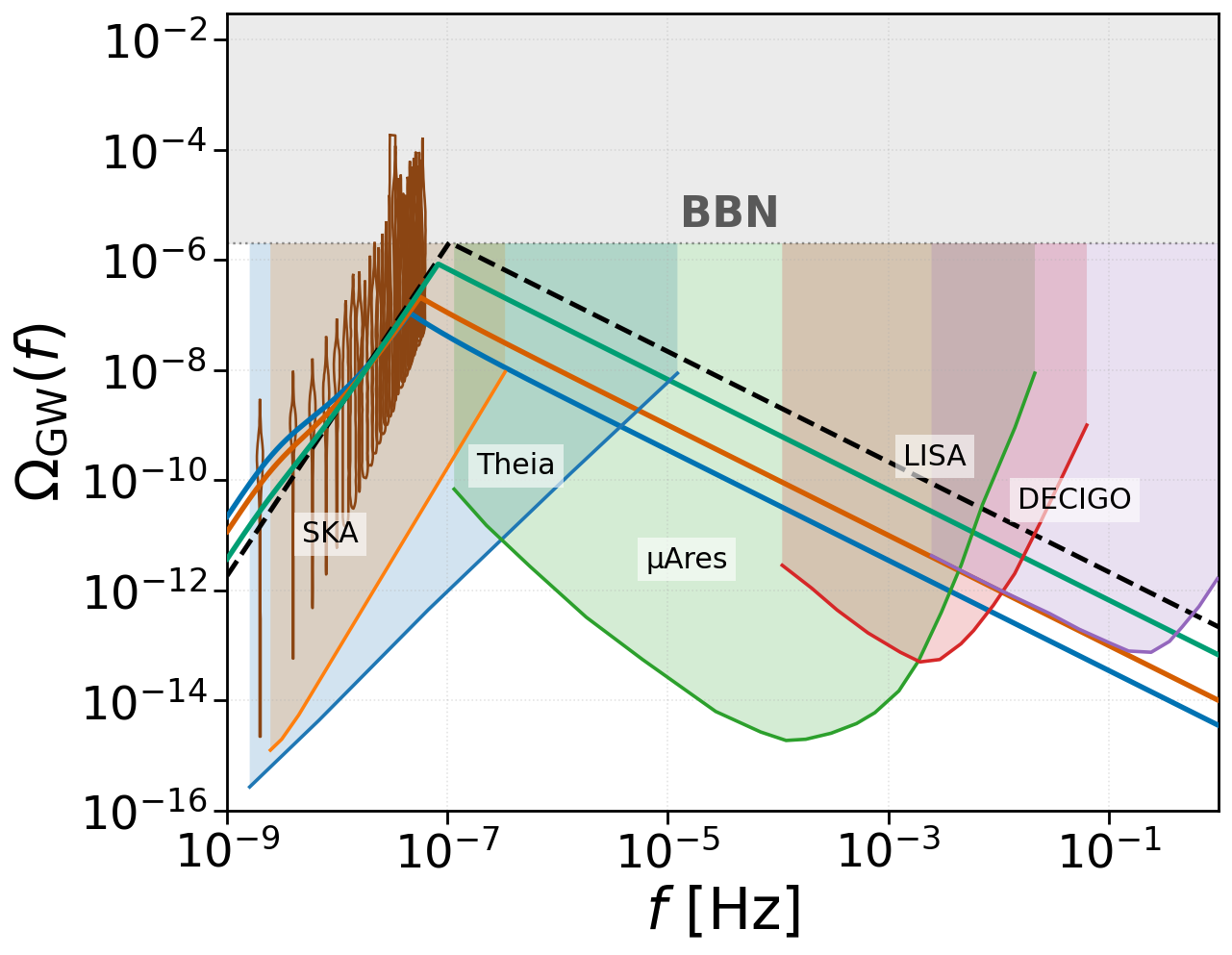}
\includegraphics[width=0.49\linewidth]{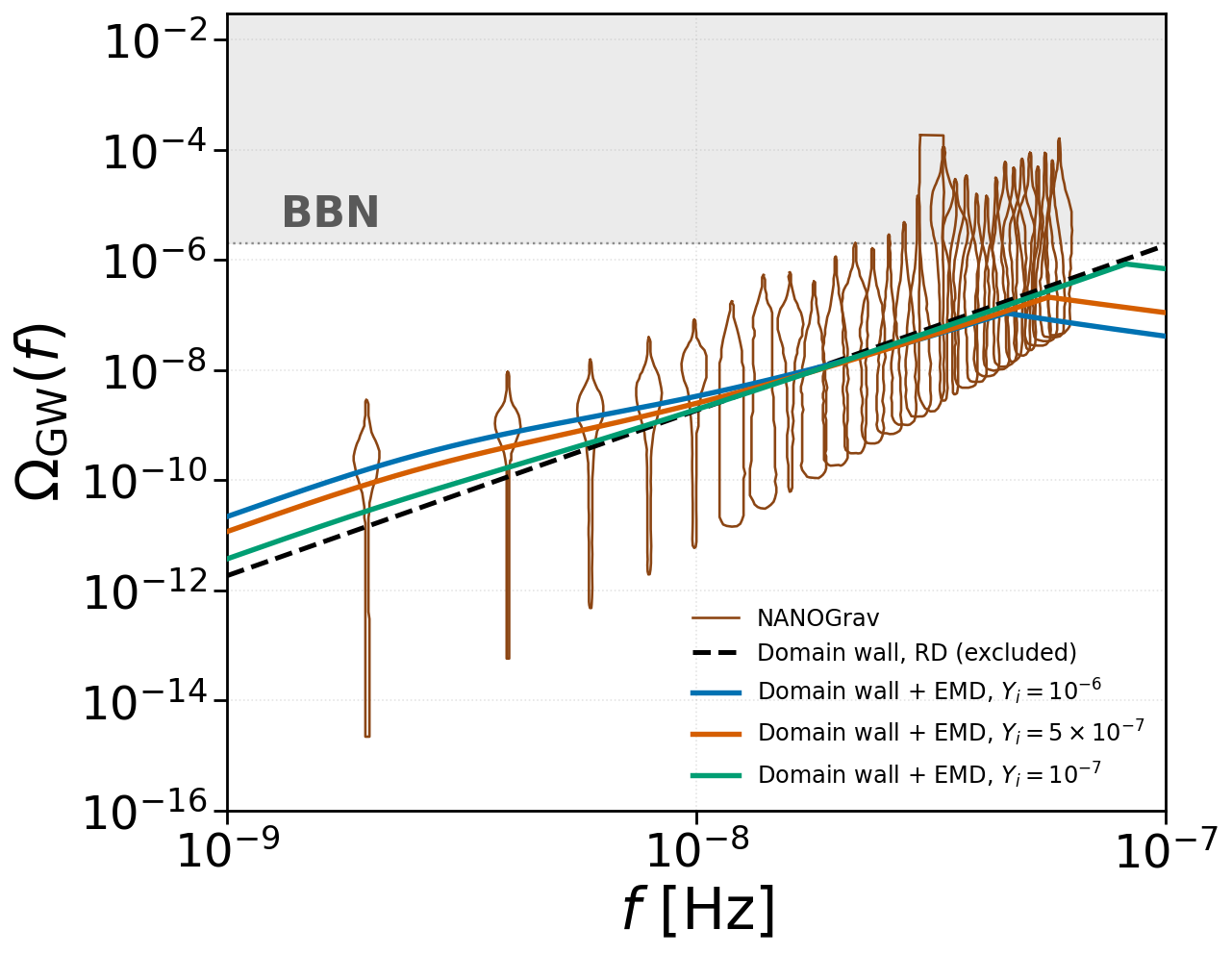} 
\caption{\it Here we show how a period of gravitino-induced early matter domination can reshape a domain-wall gravitational-wave background, with particular emphasis on the NANOGrav frequency range. The brown violins show the NANOGrav 15-year free-spectrum posteriors, while the black dashed curve gives the benchmark domain-wall spectrum during radiation domination. The coloured curves show the corresponding spectra after periods of early matter domination for a fixed gravitino mass of $m_{3/2}=1000\,\mathrm{TeV}$ and the initial abundances $Y_i$ listed in the legend. The grey shaded region is excluded by the BBN bound on the integrated gravitational-wave energy density~\cite{Clarke:2020bil}. \textbf{Left panel:} full frequency range together with the projected sensitivities of future gravitational-wave experiments. \textbf{Right panel:} zoom into the NANOGrav frequency range, highlighting the spectral modification induced by early matter domination. Further details are discussed in the main text.}
    \label{fig:benchNANO}
\end{figure}

In this subsection we show how our work can be probed by NANOGrav data. We provide a benchmark fit to the data with the benchmark of a domain wall for various periods of matter domination. We show the benchmarks in Fig \ref{fig:benchNANO}.
The spectrum scales as $f^3$ during radiation domination and as $f$ for modes that enter the horizon during an early matter-dominated era, as expected for a local source of a stochastic gravitational-wave background. We also show the BBN bound on the integrated gravitational-wave energy density from Ref.~\cite{Clarke:2020bil}.

The radiation-dominated benchmark is compared with several spectra that undergo periods of early matter domination. All benchmarks have the same gravitino mass but different initial abundances. Because the mass is fixed, the end of early matter domination occurs at the same frequency in each case. Consequently, the low-frequency part of every spectrum changes from the $f$ scaling back to the radiation-dominated $f^3$ scaling at the same point.

A larger initial abundance causes early matter domination to begin earlier and therefore last longer. This produces a larger entropy dilution, a greater shift of the spectral peak towards lower frequencies, and a stronger suppression of the gravitational-wave spectrum. We show this example as a case where EMD can save parameter space believed to be ruled out. 

\subsection{Astrophysical Foregrounds}
Our analysis assumes a primordial gravitational-wave background that carries the imprint of the gravitino-dominated epoch. Astrophysical sources, including supermassive black-hole binaries at PTA frequencies \cite{Jaffe:2002rt,Sesana:2008mz}, Galactic white-dwarf binaries in the LISA band \cite{Hils:1990vc, Farmer:2003pa}, and compact stellar binaries at higher interferometer frequencies \cite{Phinney:2001di,Farmer:2003pa}, generate gravitational waves at much later times and therefore do not acquire this imprint. These contributions can obscure the primordial features if their amplitude is comparable to or larger than the primordial background over the relevant frequency range. A conservative condition for the primordial feature to remain directly visible without foreground modelling is therefore
\begin{equation}
    \Omega_{\rm GW}^{\rm prim}(f)
    >
    \Omega_{\rm GW}^{\rm astro}(f).
    \label{eq:astro_foreground_condition}
\end{equation}
If the primordial component dominates, their impact is correspondingly reduced; otherwise, the astrophysical contribution must be accounted for through joint modelling or subtraction \cite{Adams:2013qma,Regimbau:2016ike}, with sufficient precision to distinguish the spectral break. If the astrophysical foreground is modelled and subtracted, a sufficient condition for the gravitino-induced feature to remain distinguishable is
\begin{equation}
    \delta\Omega_{\rm GW}^{\rm astro}(f)
    <
    \left|
    \Omega_{\rm GW}^{\rm prim,no\,EMD}(f)
    -
    \Omega_{\rm GW}^{\rm prim,EMD}(f)
    \right|,
    \label{eq:astro_residual_condition}
\end{equation}
where $\delta\Omega_{\rm GW}^{\rm astro}(f)$ denotes the residual uncertainty in the astrophysical foreground after modelling or subtraction. These conditions provide simple conservative criteria for the feature to remain distinguishable and do not require specifying a detailed astrophysical foreground
model. A quantitative treatment of specific astrophysical foreground populations,
their modelling, and their subtraction is beyond the scope of the present work. However the impact of it is expected to be model independent, and it has been studied in other models, 
to which we refer the interested reader ( see for example the discussion and figure in section 4.1 of ref.~\cite{Fu:2024rsm}). 
Once the feature is resolved, its frequency retains the direct connection to the
gravitino mass established above. 

\subsection{Non-Transient and Sourced GWBs}

We now briefly consider gravitational-wave backgrounds that remain actively
sourced during the early matter-dominated epoch. Following the treatment of
previous work, the gravitational-wave energy-density equation becomes
\begin{equation}
\dot{\rho}_{\rm GW}(t,f)
+ 4H(t)\rho_{\rm GW}(t,f)
=
\mathcal{S}(t,f),
\label{eq:sourced_gw_fluid}
\end{equation}
where $\mathcal{S}(t,f)$ denotes the gravitational-wave source term. The
evolution of the fractional gravitational-wave energy density is then
\begin{equation}
\frac{d\Omega_{\rm GW}(a,f)}{d\ln a}
=
\left(3w(a)-1\right)\Omega_{\rm GW}(a,f)
+
\frac{\mathcal{S}(a,f)}
{H(a)\rho_{\rm tot}(a)}.
\label{eq:Omega_sourced_lna}
\end{equation}
The resulting spectrum therefore depends on both the time-dependent source and the cosmological transfer function. Importantly, the lower characteristic frequency $f_1$ remains fixed by the
end of early matter domination,
\begin{equation}
f_1^{\rm sourced}=f_1^{\rm transient}=\frac{a_{\rm end}H_{\rm end}}{2\pi a_0},
\label{eq:f1_sourced}
\end{equation}
and hence the relation between $f_1$ and the decay rate remains unchanged,
\begin{equation}
f_1=20.6\left(\frac{\Gamma}{\mathrm{GeV}}\right)^{1/2}
\mathrm{Hz}.
\label{eq:f1_sourced_Gamma}
\end{equation}
Although the source-dependent spectral structure may make this feature more difficult to identify observationally, its underlying cosmological scale is unchanged. By contrast, the higher-frequency feature depends on the detailed evolution
of the gravitational-wave source, and in general $f_2^{\rm sourced} \neq f_2^{\rm transient}.$
For the gravitino scenario considered here, the end of early matter domination
is determined by gravitino decay, which is fixed by the gravitino mass, a measurement of $f_1$
can therefore determine both $\Gamma_{3/2}$ and $m_{3/2}$. Thus, the gravitino
mass can be inferred not only from a freely propagating gravitational-wave background, but also from a continuously sourced background, provided that the $f_1$ feature can be identified \cite{Ghoshal:2026ros}. However, the extraction of the initial abundance then depends on the source. This demonstrates the novelty of the gravitino dominated scenario as a Lagrangian parameter can be directly extracted from generic primordial gravitational wave backgrounds. 
\section{Discussion \& Conclusion}\label{sec:conclusion}
\noindent Supersymmetry, together with its local extension supergravity, remains one of the most compelling frameworks for physics beyond the Standard Model. In particular, supergravity theories, and therefore string theories, generically predict the existence of the gravitino: the spin-$3/2$ superpartner of the graviton.\\ The cosmological presence of gravitinos in the early universe gives rise to the well-known gravitino problem, since long-lived gravitinos can disrupt the successful predictions of Big Bang Nucleosynthesis (BBN) through their late-time decay products. A simple and widely studied resolution is to consider sufficiently heavy gravitinos, with masses $m_{3/2} \gtrsim \mathcal{O}(100)\,\text{TeV}$, such that they decay before the onset of BBN. However, gravitinos at these mass scales are far beyond the direct reach of present and near-future collider experiments, motivating the search for alternative cosmological probes of heavy supersymmetric sectors.\\
In this work, we have shown that gravitinos in this regime can nevertheless be probed through their impact on the stochastic gravitational wave background. Despite decaying before BBN, gravitinos naturally generate a period of early matter domination, leaving a characteristic imprint on the spectrum in the form of two frequencies, $f_1$ and $f_2$. These features provide a direct mapping between gravitational wave observables and the underlying particle physics parameters. \\
In previous analyses \cite{Datta:2025vyu, Ghoshal:2026ros}, the mapping between a long-lived particle and GWB observables involved the initial abundance, mass, and decay rate as inputs. By specifying the gravitino as the dominant particle, this analysis is refined, since the parameter space reduces to a direct mapping between the gravitino parameters (initial abundance and the gravitino mass) to the two GWB observables, $f_1$ and $f_2$ as shown in Fig \ref{fig:GWB_inference_chain}. In particular, the frequency $f_1$ alone determines the gravitino mass, while the combination of $f_1$ and $f_2$ allows for the extraction of both the mass and initial abundance. In addition, the gravitino naturally satisfies Eq \ref{eq:Tcondition}.\\
Future gravitational wave observatories span an enormous frequency range, allowing sensitivity to gravitino masses from $\mathcal{O}(100)$ TeV up to $\mathcal{O}(10^{10})$ TeV. In particular, pulsar timing array experiments such as SKA probe the lower end of this range, space-based interferometers such as LISA extend sensitivity to intermediate scales, and ground-based detectors such as ET and Cosmic Explorer probe the highest masses. These ranges cover many orders of magnitude, from the scale required to solve the gravitino problem up to scales approaching those associated with inflation.\\
Gravitational wave backgrounds therefore provide a powerful and complementary probe of supergravity. In particular, they offer a direct window onto the gravitino sector at mass scales far beyond those accessible to collider experiments, opening a new avenue to test high-scale supergravity in the early universe.
\section*{Acknowledgements}
Authors thank Graham White and Anish Ghoshal for discussions on Gravitational Wave Detectors as well as for comments on the manuscript. AS and SFK acknowledge the STFC Consolidated Grant ST/X000583/1. SFK is funded by a Leverhulme Trust Emeritus Fellowship Grant. AS thanks the University of Southampton School of Physics and Astronomy for the support of a Mayflower PhD scholarship.
\bibliographystyle{apsrev4-1}

\bibliography{ref}
\end{document}